\documentclass[twocolumn]{aastex631}

\usepackage{bm}

\shorttitle{Pulsar timing to spherical GWs}
\shortauthors{Kubo et al.}

\begin{document}

\title{
Pulsar Timing  Response to Gravitational Waves with Spherical Wavefronts 
from a Massive Compact Source in the Quadrupole Approximation
}

\author{Ryousuke Kubo}
\affiliation{
Graduate School of Science and Technology, \\
Hirosaki University, \\
3 Bunkyo-cho \\
Hirosaki, Aomori 036-8055, Japan}

\author{Kakeru Yamahira}
\affiliation{
Graduate School of Science and Technology, \\
Hirosaki University, \\
3 Bunkyo-cho \\
Hirosaki, Aomori 036-8055, Japan}

\author[0000-0001-9442-6050]{Hideki Asada}
\affiliation{
Graduate School of Science and Technology, \\
Hirosaki University, \\
3 Bunkyo-cho \\
Hirosaki, Aomori 036-8055, Japan}



\begin{abstract}
Pulsar timing arrays (PTAs) are searching for nanohertz-frequency gravitational waves (GWs) through cross-correlation of pulse arrival times from a set of radio pulsars. 
PTAs have relied upon a frequency-shift formula of the pulse, 
where planar GWs are usually assumed. 
Phase corrections due to the wavefront curvature 
have been recently discussed.  
In this paper, 
frequency-shift and timing-residual formulae are 
derived for GWs with fully spherical wavefronts 
from a compact source such as a binary of supermassive black holes, 
where the differences 
in the GW amplitude and direction between the Earth and the pulsar 
are examined in the quadrupole approximation. 
By using the new formulae, 
effects beyond the plane-wave approximation are discussed, 
and a galactic-center PTA as well as 
nearby GW source candidates  
are also mentioned. 
\end{abstract}

\keywords{Gravitational waves  (678); 
Pulsar timing method (1305); 
Gravitational wave astronomy (675); 
Millisecond pulsars (1062); 
Supermassive black hole (1663)}


\section{Introduction} \label{sec:intro}
The method of using radio pulse timing to search for gravitational waves (GWs) 
can be dated back to 
\citep{Estabrook1975,Sazhin1978,Detweiler1979,Hellings1983}.
A possible deviation from the expected noise 
has been reported by the NANOGrav team 
 \citep{Arzoumanian2020,Antoniadis2022}, 
and has been argued 
by several teams  
\citep{Pol2021,Alam2021a,Alam2021b,Arzoumanian2021a,Arzoumanian2021b,Kaiser2022,Goncharov2022}.  
It is expected that the first detection by the International Pulsar Timing Array  consortium may come soon 
\citep{Castelvecchi2022}.

PTA studies have relied upon a frequency-shift formula of a radio pulse, 
where planar GWs are usually assumed 
\citep{Estabrook1975,Detweiler1979,Hellings1983}. 
The wavefront curvature for a distant GW source has been 
discussed as a correction; 
the Fresnel approximation is discussed 
\citep{Deng2011,McGrath2021}.   
Toward PTA cosmology, 
the importance of distinguishing the comoving distance 
from the luminosity distance  
has been examined 
\citep{D'Orazio2021,McGrath2022}. 

One may ask how a compact GW source  
affects PTA observational signatures. 
The main purpose of this paper is to discuss 
a PTA detector response to GWs from a compact source 
such as binary supermassive black holes (SMBHs), 
which are thought to harbor in galactic centers. 
In section 2, we examine 
the pulse's frequency shift and timing residual 
to GWs with fully spherical wavefronts 
from a compact source. 
In section 3, 
possible effects beyond the plane-wave approximation 
are discussed. 
Section 4 summarizes this paper. 
Throughout this paper, $c=1$ and the Latin indices $i, j$ 
run from 1 to 3.

\section{PTA response: From a planar wave to a spherical wave}
\subsection{PTA response to GWs}
We begin with a derivation of the 
pulse's frequency shift  
\citep{Creighton2013,Maggiore2018}. 
In particular, 
we do not assume planar GWs such that our result can be applied 
also to 
GWs with spherical wavefronts as shown 
in next subsection.

We suppose that a radio pulse is emitted by a pulsar (P) 
at time $t_P$ and arrives at the Earth (E) at $t_E$ 
and position $\bm{x}_E$. 
The radio signal obeys the null condition as 
\begin{equation}
0 = -dt^2 + (\delta_{ij} + h_{ij}^{TT} ) dx^i dx^j ,
\label{null}
\end{equation}
where the transverse and traceless (TT) gauge is used 
and $h_{ij}^{TT}$ is GW perturbations. 
The unit vector along the pulse is 
$dx^i / d\ell = - n_P^i$, where $\ell$ denotes the spatial length 
and $n_P^i$ denotes the unit vector from E to P. 

Eq. (\ref{null}) is rearranged as 
\begin{equation}
d\ell = 
\left( 
1 - \frac12 n_P^i n_P^j h_{ij}^{TT} (t, \bm{x}(t)) 
\right) 
dt + O(h^2) ,
\label{dell}
\end{equation}
where $O(h^2)$ denotes  the second order terms 
in $h_{ij}^{TT}$.  
In the TT gauge, the spatial coordinates of E and P 
are constants and the clocks on them are also aligned. 
See e.g. \citet{Creighton2009} for the role of the gauge in 
pulsar timing experiments.

The distance $L$ between E and P is 
\begin{eqnarray}
L &=& \int_P^E d\ell 
\nonumber\\
&=&
t_E - t_P 
- \frac12 n_P^i n_P^j 
\int_{t_P}^{t_E} dt' 
h_{ij}^{TT} (t', \bm{x}(t')) 
\nonumber\\
&&+ O(h^2) . 
\label{L}
\end{eqnarray}
It follows that $t_E = t_P + L  + O(h)$. 

The spatial position of the radio signal can be written 
at the lowest order as 
\begin{equation}
\bm{x}(t) = \bm{x}_E + (t_P + L - t) \bm{n}_P + O(h) . 
\label{x}
\end{equation}
Substituting Eq. (\ref{x}) into Eq. (\ref{L}) leads to 
\begin{eqnarray}
L 
&=& 
t_E - t_P 
\nonumber\\
&&- \frac12 n_P^i n_P^j 
\int_{t_P}^{t_P + L} dt' 
h_{ij}^{TT} (t', \bm{x}_E + (t_P - t' +L) \bm{n}_P) 
\nonumber\\
&&+ O(h^2) . 
\label{L2}
\end{eqnarray}
This agrees with e.g. Eq. (23.5) in \citet{Maggiore2018}. 

For a radio pulse emitted at $t_{em}$ and observed at $t_{obs}$,  
the notation change in Eq. (\ref{L2}) 
as $t_P \to t_{em}$ and $t_E \to t_{obs}$ leads to 
\begin{eqnarray}
L 
&=& 
t_{obs}- t_{em} 
\nonumber\\
&&- \frac12 n_P^i n_P^j 
\int_{t_{em}}^{t_{em} + L} dt' 
h_{ij}^{TT} (t', \bm{x}_{E} + (t_{em} - t' +L) \bm{n}_P) 
\nonumber\\
&&
+ O(h^2) .
\label{L-obs1}
\end{eqnarray}
For the next pulse emitted at $t_{em}'$ and observed at $t_{obs}'$, 
\begin{eqnarray}
L 
&=& 
t_{obs}'- t_{em}' 
\nonumber\\
&&
- \frac12 n_P^i n_P^j 
\int_{t_{em}'}^{t_{em}' + L} dt' 
h_{ij}^{TT} (t' + T_P, \bm{x}_{E} + (t_{em}- t' +L) \bm{n}_P) 
\nonumber\\
&&
+ O(h^2) . 
\label{L-obs2}
\end{eqnarray}
The linear perturbation by GWs suffices in the scope of this paper. 
Hence, $O(h^2)$ is omitted in the rest of this paper. 

The observed period and intrinsic one of the radio pulse are 
$T_E = t_{obs}' - t_{obs}$ 
and 
$T_P = t_{em}' - t_{em}$, respectively. 
In the TT gauge, Eqs. (\ref{L-obs1}) and (\ref{L-obs2})
have the same separation $L$. 
Thereby, 
the deviation of the observed period from the intrinsic one is obtained as 
\begin{eqnarray}
\Delta T &\equiv& T_E - T_P 
\nonumber\\
&=& 
\frac12 T_p n_P^i n_P^j 
\nonumber\\
&&
\times\int_{t_{em}'}^{t_{em}' + L} dt' 
\left[\frac{\partial}{\partial t'} 
h_{ij}^{TT} (t', \bm{x})\right]_{\bm{x}=\bm{x}_E + (t_{em} - t' +L) \bm{n}_P} , 
\nonumber\\
&&
\label{DeltaT} 
\end{eqnarray} 
where the GW period $T_{GW} \gg T_P \sim 1$ msec. is used.

The redshift due to the pulse period shift becomes  
\begin{eqnarray}
z &\equiv& \frac{\Delta T}{T_P}
\nonumber\\
&=& 
\frac12 n_P^i n_P^j 
\int_{t_{em}'}^{t_{em}' + L} dt' 
\left[\frac{\partial}{\partial t'} 
h_{ij}^{TT} (t', \bm{x})\right]_{\bm{x}=\bm{x}_E + (t_{em} - t' +L) \bm{n}_P} .
\nonumber\\
&&
\label{Doppler} 
\end{eqnarray} 
This causes the frequency shift of the radio signal as 
$\Delta f/f_E = -z$ 
because $f_E=1/T_E$, $f_P=1/T_P$ and $\Delta f \equiv f_E - f_P$. 
The partial differentiation $\partial/\partial t'$ acts only on the time argument 
in $h_{ij}^{TT}$ but not on the spatial argument. 
Therefore, the integrand in Eq. (\ref{Doppler}) cannot be recast 
into a total differentiation, 
except for planar GWs. 
In a general situation without any approximation, therefore, 
we need perform the integral along the radio path.

\subsection{Frequency shift by spherical wavefronts}
\begin{figure}[]
\plotone{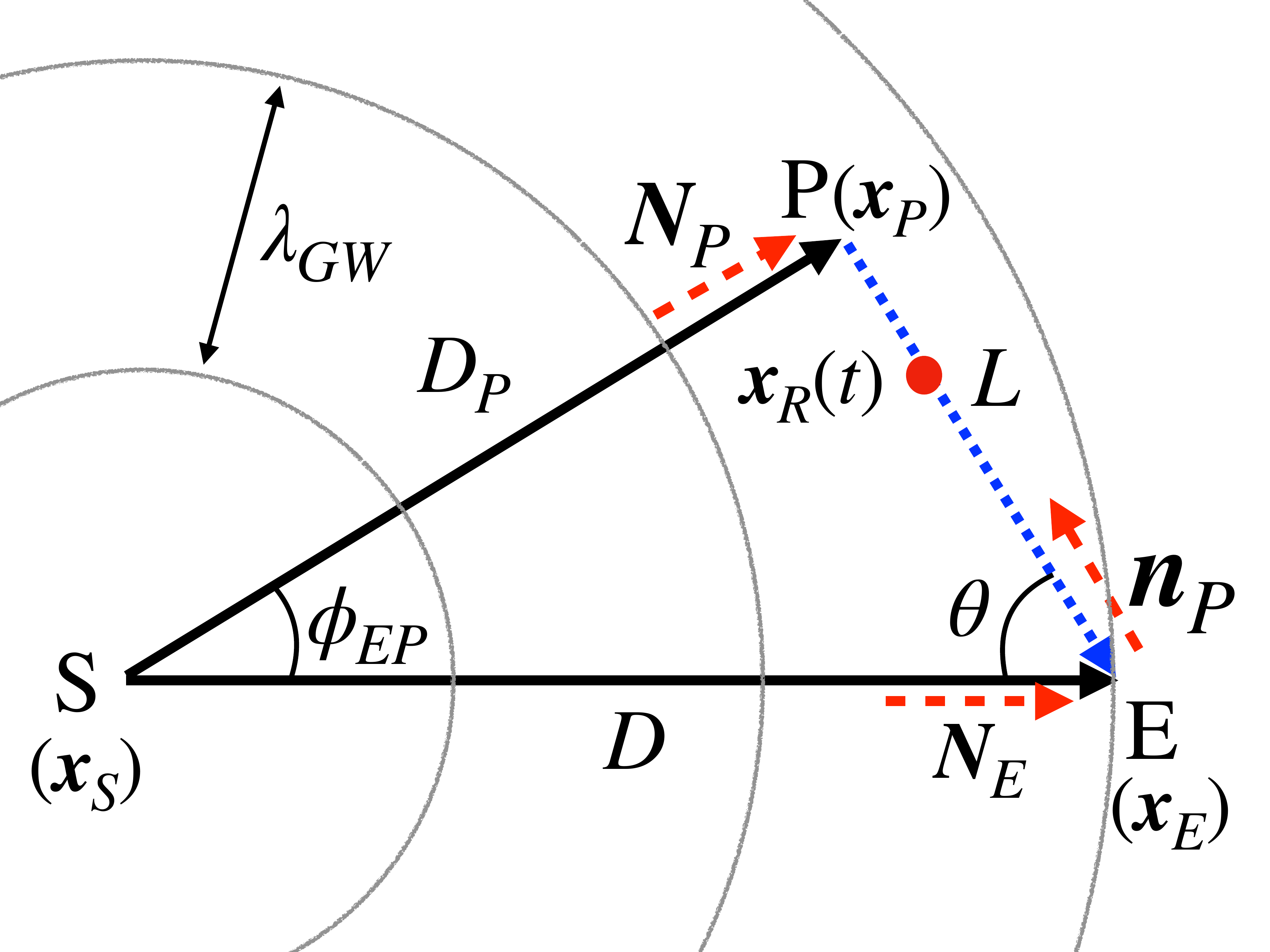}
\caption{Configuration of 
the Earth (E), a pulsar (P) and a GW source (S). 
The black solid arrows denote GW paths to E or P, 
where 
red (in color) dashed arrows indicate 
the unit vectors along the GW propagation, 
$\bm{N}_E$ and $\bm{N}_P$. 
The blue (in color) dotted arrow means the radio signal, 
where the unit vector from E to P is 
a red (in color) arrow $\bm{n}_P$. 
The distances between E and S,  
between E and P, 
and between P and S are 
$D$, $L$ and $D_P$, respectively. 
}
\label{fig:config}
\end{figure}

Figure \ref{fig:config} shows 
a configuration of the Earth, a pulsar and a GW source (S). 
We suppose $\lambda_P \ll \lambda_{GW} < D \sim D_P$, 
where $\lambda_P$ and $\lambda_{GW}$ are wavelengths 
of the radio pulse and the GW, respectively.

In the quadrupole approximation, 
the GW at a radio pulse position $\bm{x}_R(t)$ 
is expressed as 
\begin{equation}
h_{ij}^{TT} (t, \bm{x}_R(t)) 
= 
\frac{q_{ij}^{TT}(U, \bm{N}_R(t) )}{|\bm{x}_R(t) - \bm{x}_S|} ,
\label{h}
\end{equation}
where $q_{ij}^{TT}$ denotes the radiative part 
(the TT part of the second time derivative of 
the mass quadrupole moment multiplied by $2G$) 
of the GW source at $\bm{x}_S$ 
and 
$U \equiv t- |\bm{x}_R(t) - \bm{x}_S|$ is the retarded time.

We should note our treatment of the radio pulse position. 
In Eq. (\ref{h}), $q_{ij}^{TT}$ in its numerator causes the linear perturbation. 
In the linear approximation of Eq. (\ref{h}), 
therefore, it suffices to use $\bm{x}_R(t) \equiv \bm{x}_E + (t_P + L - t) \bm{n}_P$ 
by safely ignoring $O(h)$ in Eq. (\ref{x}).

The GW propagation direction at $R$ is 
\begin{equation}
\bm{N}_R(t) \equiv 
\frac{\bm{x}_R(t) - \bm{x}_S}{|\bm{x}_R(t) - \bm{x}_S|} .
\label{N}
\end{equation}
Note that $q_{ij}^{TT}$ depends on $\bm{N}_R(t)$ 
via the TT projection operator, 
where 
$\bm{N}_R(t)$ as a function of time  
causes a deviation from a plane-wave case.

From Eq. (\ref{h}), we obtain 
\begin{eqnarray}
&&
\left[
\frac{\partial}{\partial t}  
h_{ij}^{TT} (t, \bm{x}) 
\right]_{\bm{x} = \bm{x}_R(t)}
\nonumber\\
&=&
\frac{1}{|\bm{x}_R(t) - \bm{x}_S|} 
\left[
\frac{\partial q_{ij}^{TT}(U, \bm{N})}{\partial U}
\right]_{\bm{N} = \bm{N}_R(t)} .
\label{doth}
\end{eqnarray}
We shall examine subtle calculations of the right-hand side of 
Eq. (\ref{doth}).

By direct calculations, 
we obtain 
\begin{eqnarray}
&&
\frac{d}{dt} 
q_{ij}^{TT} (U, \bm{N}_R(t)) 
\nonumber\\
&=&
\frac{dU}{dt}
\left[
\frac{\partial q_{ij}^{TT}(U, \bm{N})}{\partial U} 
\right]_{\bm{N} = \bm{N}_R(t)}
\nonumber\\
&&
+ 
\frac{d \bm{N}_R(t)}{dt}
\left[
\frac{\partial q_{ij}^{TT}(U, \bm{N})}{\partial \bm{N}} 
\right]_{\bm{N} = \bm{N}_R(t)} , 
\label{dq}
\end{eqnarray}
of which each term is calculated separately below. 

First, we obtain 
\begin{eqnarray}
\frac{dU}{dt} 
&=&
\frac{d}{dt} 
(t - |\bm{x}_R(t) - \bm{x}_S|) 
\nonumber\\
&=& 
1 - \frac{d\bm{x}_R(t)}{dt} 
\cdot 
\frac{\bm{x}_R(t) - \bm{x}_S}{|\bm{x}_R(t) - \bm{x}_S|} 
\nonumber\\
&=& 
1 + \bm{n}_P\cdot\bm{N}_R(t) , 
\label{part1}
\end{eqnarray}
where
$d\bm{x}_R(t)/dt = - \bm{n}_P$ 
available from the time derivative of Eq. (\ref{x}) 
is used 
in the third line. 

Next, we find 
\begin{eqnarray}
\left[
\frac{\partial q_{ij}^{TT}(U, \bm{N})}{\partial U} 
\right]_{\bm{N} = \bm{N}_R(t)}
= O\left(\frac{q}{\lambda_{GW}}\right) , 
\label{part2}
\end{eqnarray}
where $q$ denotes the magnitude of $|q_{ij}^{TT}(U, \bm{N}_R(t))|$. 

Thirdly, 
\begin{eqnarray}
\frac{d \bm{N}_R(t)}{dt} 
&=& 
\frac{d}{dt} 
\frac{\bm{x}_R(t) - \bm{x}_S}{|\bm{x}_R(t) - \bm{x}_S|}
\nonumber\\
&=& O\left( \frac1D \right) ,
\label{part3} 
\end{eqnarray}
where we use 
$d \bm{x}_R(t)/dt = -\bm{n}_P = O(1)$ 
and 
$|\bm{x}_R(t) - \bm{x}_S| = O(D)$. 

Finally, we obtain 
\begin{eqnarray}
\left[
\frac{\partial q_{ij}^{TT}(U, \bm{N})}{\partial \bm{N}} 
\right]_{\bm{N} = \bm{N}_R(t)}
= O(q) . 
\label{part4} 
\end{eqnarray}

From Eqs. (\ref{part1})-(\ref{part4}), 
the second term in the right-hand side of Eq. (\ref{dq}) 
is smaller by factor of $O(\lambda_{GW}/D)$ than the first term. 
Therefore, 
Eq. (\ref{dq}) 
is rearranged as 
\begin{eqnarray} 
&&
\frac{d}{dt} 
q_{ij}^{TT} (U, \bm{N}_R(t)) 
\nonumber\\
&=&
(1 + \bm{n}_P\cdot\bm{N}_R(t)) 
\left[
\frac{\partial q_{ij}^{TT}(U, \bm{N})}{\partial U} 
\right]_{\bm{N} = \bm{N}_R(t)}
\left[
1 + O\left( \frac{\lambda_{GW}}{D} \right)
\right] .
\nonumber\\
&&
\label{hint1}
\end{eqnarray}

From Eq. (\ref{h}), we obtain 
\begin{eqnarray}
\frac{d}{dt}  
h_{ij}^{TT} (t, \bm{x}_R(t))
&=& 
\frac{1}{|\bm{x}_R(t) - \bm{x}_S|} 
\frac{d}{dt} 
q_{ij}^{TT}(U, \bm{N}_R(t)) 
\nonumber\\
&&
+ 
q_{ij}^{TT}(U, \bm{N}_R(t)) 
\frac{d}{dt} 
\frac{1}{|\bm{x}_R(t) - \bm{x}_S|} .  
\nonumber\\
\label{dh} 
\end{eqnarray}
Here, we obtain 
\begin{equation}
\frac{d}{dt} 
\frac{1}{|\bm{x}_R(t) - \bm{x}_S|} 
= O\left(\frac{1}{D^2}\right) ,
\label{hint2}
\end{equation}
where $d \bm{x}_R(t)/dt = O(1)$ 
and 
$|\bm{x}_R(t) - \bm{x}_S| = O(D)$ are used again.

Note that a pulse trajectory is perturbed by GWs 
\citep{Finn}, but the perturbation 
can induce terms of $O(h/D)$ in Eq. (\ref{hint2}), 
which cause only  $O(h^2)$ in Eq. (\ref{dh}). 
Therefore, the perturbed trajectory can be ignored in the present paper.

From Eqs. (\ref{part2}), (\ref{hint1}) and (\ref{hint2}), 
the first term and second one in the right-hand side 
of Eq. (\ref{dh}) are 
$O(q/(D\lambda_{GW}))$ and $O(q/D^2)$, respectively. 
Namely, the first term is larger 
by factor of $O(D/\lambda_{GW})$ 
than the second one. 
We thus find 
\begin{eqnarray}
&&
\frac{d}{dt}  
h_{ij}^{TT} (t, \bm{x}_R(t)) 
\nonumber\\
&=&
\frac{1}{|\bm{x}_R(t) - \bm{x}_S|} 
\frac{d}{dt} 
q_{ij}^{TT}(U, \bm{N}_R(t)) 
\left[
1 + O\left( \frac{\lambda_{GW}}{D} \right)
\right] 
\nonumber\\
&=& 
\frac{1 + \bm{n}_P\cdot\bm{N}_R(t)}{|\bm{x}_R(t) - \bm{x}_S|} 
\left[
\frac{\partial q_{ij}^{TT}(U, \bm{N})}{\partial U} 
\right]_{\bm{N} = \bm{N}_R(t)}
\left[
1 + O\left( \frac{\lambda_{GW}}{D} \right)
\right] ,
\nonumber\\
\label{dhdt}
\end{eqnarray}
where Eq. (\ref{hint1}) is used in the third line.

From Eqs. (\ref{doth}) and (\ref{dhdt}) , we obtain 
\begin{eqnarray}
&&
\frac{\partial}{\partial t}  
h_{ij}^{TT} (t, \bm{x}_R(t)) 
\nonumber\\
=
&&
\frac{1}{1 + \bm{n}_P\cdot\bm{N}_R(t)} 
\frac{d}{dt}
h_{ij}^{TT} (t, \bm{x}_R(t)) 
\left[
1 + O\left( \frac{\lambda_{GW}}{D} \right)
\right]
\nonumber\\
=
&&
\frac{d}{dt}
\left(
\frac{1}{1 + \bm{n}_P\cdot\bm{N}_R(t)} 
h_{ij}^{TT} (t, \bm{x}_R(t)) 
\right)
\left[
1 + O\left( \frac{\lambda_{GW}}{D} \right)
\right] ,
\nonumber\\
\label{hint3}
\end{eqnarray}
where  
we use in the last line 
\begin{equation}
\frac{d}{dt} 
\frac{1}{1 + \bm{n}_P \cdot \bm{N}_R(t)}
= 
O\left(\frac{1}{D}\right) , 
\label{hint4}
\end{equation}
and $d( h_{ij}^{TT})/dt =O(h/\lambda_{GW})$.


Substituting Eq. (\ref{hint3}) into Eq. (\ref{Doppler}) 
leads to 
\begin{eqnarray}
z 
&=& 
\frac12 n_P^i n_P^j 
\nonumber\\
&&
\times
\int_{t_{em}'}^{t_{em}' + L} dt' 
\frac{d}{dt'} 
\left(
\frac{1}{1 + \bm{n}_P\cdot\bm{N}_R(t')} 
h_{ij}^{TT} (t', \bm{x}_R(t')) 
\right)
\nonumber\\
&&
\times 
\left[
1 + O\left( \frac{\lambda_{GW}}{D} \right)
\right]  
\nonumber\\
&=&
\frac12 n_P^i n_P^j 
\left[
\frac{1}{1 + \bm{n}_P\cdot\bm{N}_E}
h_{ij}^{TT} (t_E, \bm{x}_E) 
\right.
\nonumber\\
&&~~~~~~~~~~~
\left.
- \frac{1}{1 + \bm{n}_P\cdot\bm{N}_P}
h_{ij}^{TT} (t_P, \bm{x}_P) 
\right]  
\nonumber\\
&&
+ O\left( \frac{h \lambda_{GW}}{D}\right) ,
\label{Doppler2} 
\end{eqnarray} 
where the remainder term is 
$\sim 10^{-7} (\lambda_{GW}/10 \mbox{pc})(100 \mbox{Mpc}/D)\times O(h)$ 
and hence it can be safely ignored. 
The plane-wave formula is recovered by Eq. (\ref{Doppler2}) 
in the limit $L/D \to 0$ for which $\bm{N}_P \to \bm{N}_E$. 
Because of the retardation in Eq. (\ref{h}), 
$h_{ij}^{TT} (t_E, \bm{x}_E)$ and $h_{ij}^{TT} (t_P, \bm{x}_P)$ 
come from the quadrupole moments 
at the GW source time
$t_E - D$ and $t_P - D_P$, respectively, 
where $t_P = t_E-L$. 

If we assumed $\bm{N}_E = \bm{N}_P$, $D = D_P$ and $\lambda_{GW} < L$, 
Eq. (\ref{Doppler2}) could recover the wavefront-curvature effects 
 in the literature 
 \citep{Deng2011,McGrath2021,D'Orazio2021}, 
where $\bm{N}_E = \bm{N}_P$ and $D = D_P$ are 
approximations for $L \ll D$, 
because $|\bm{N}_E - \bm{N}_P| = O(L/D)$.

\begin{figure}[]
\plotone{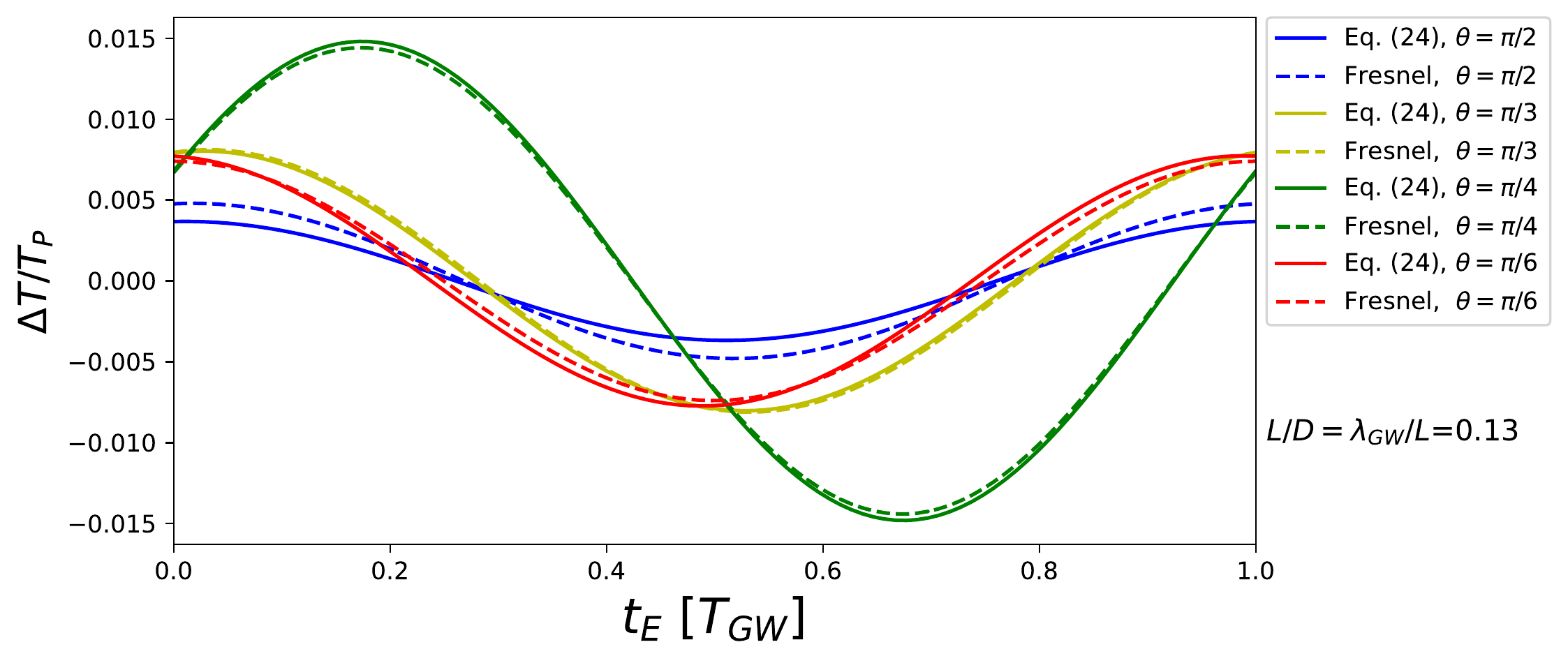}
\caption{
Redshift 
in PTA due to GWs from an edge-on circular binary. 
The vertical axis denotes the 
redshift 
$z = \Delta T/T_P$. 
The horizontal axis denotes $t_E$ in the unit of $T_{GW}$.  
The colored dotted curves include only the phase correction 
by the Fresnel model \citep{McGrath2021}. 
The colored solid curves 
take account of also the GW amplitude and direction corrections 
via Eq. (\ref{Doppler2}) . 
For the curves to be recognized by eye, 
$L/D = \lambda_{GW}/L = 0.13$ is chosen. 
}
\label{fig:plot1}
\end{figure}

\subsection{Pulsar timing residual by spherical wavefronts}
Finally, we mention the pulsar timing residual induced by the GWs, 
which is the integrated fractional period shift over the observation time 
\citep{Creighton2013,Maggiore2018,McGrath2021}. 
The timing residual is 
\begin{eqnarray}
\mbox{Res}(t_{obs}) = \int^{t_{obs}}_0 dt_E \frac{\Delta T}{T_P} , 
\label{Res}
\end{eqnarray}
where the initial time of the observation of interest is chosen as 
$t_E = 0$ without loss of generality 
and the observation time is $t_E = t_{obs}$. 

By substituting Eq. (\ref{Doppler2}) into Eq. (\ref{Res}), 
we obtain 
\begin{eqnarray}
&&\mbox{Res}(t_{obs})
\nonumber\\
=&&
\frac12 n_P^i n_P^j 
\left[
\frac{1}{1 + \bm{n}_P\cdot\bm{N}_E}
\int^{t_{obs}}_0 dt_E h_{ij}^{TT} (t_E, \bm{x}_E) 
\right.
\nonumber\\
&&~~~~~~~~~~~~
\left.
- \frac{1}{1 + \bm{n}_P\cdot\bm{N}_P}
\int^{t_{obs}}_0 dt_E h_{ij}^{TT} (t_E-L, \bm{x}_P) 
\right]  
\nonumber\\
&&
+ O\left(\frac{h (\lambda_{GW})^2}{D}\right) ,
\label{Res2}
\end{eqnarray}
where $t_P = t_E - L$ is used in the third line, 
and 
$\int dt O(h \lambda_{GW}/D) 
\sim \int dt O(q_{ij}^{TT} \lambda_{GW}/D^2)  
\sim O(q_{ij}^{TT} (\lambda_{GW})^2/D^2) 
\sim O(h (\lambda_{GW})^2 /D)$ 
(due to $\int dt q_{ij}^{TT} \sim \lambda_{GW} q_{ij}^{TT}$)
is used in the last line. 

The integral in the timing residual is dependent 
strongly on the GW waveform of concern. 
It cannot be always reduced to a compact form. 

For its simplicity, 
let us consider a monochromatic GW regime as 
\begin{eqnarray}
h_{ij}^{TT}(t, \bm{x}) 
= 
A_{ij}^{TT}(\bm{x}) \exp\left[i\left(\frac{2\pi U}{\lambda_{GW}}\right)\right] ,
\label{monochromatic}
\end{eqnarray} 
where the GW chirp is ignored.   
Then, the timing residual is expressed compactly as 
\begin{eqnarray}
&&
\mbox{Res}(t_{obs})
\nonumber\\
&=&
\frac{i \lambda_{GW}n_P^i n_P^j}{4\pi}  
\nonumber\\
&&
\times
\left[
\frac{A_{ij}^{TT}(\bm{x}_E)}{1 + \bm{n}_P\cdot\bm{N}_E}
\left\{
\exp\left[-i\left(\frac{2\pi D}{\lambda_{GW}}\right)\right] 
\right.
\right.
\nonumber\\
&&~~~~~~~~~~~~~~~~~~~~~~~
\left.
\left.
- \exp\left[i\left(\frac{2\pi (t_{obs} - D)}{\lambda_{GW}}\right)\right] 
\right\}
\right.
\nonumber\\
&&~~~~~
\left.
- \frac{A_{ij}^{TT}(\bm{x}_P)}{1 + \bm{n}_P\cdot\bm{N}_P}
\left\{
\exp\left[-i\left(\frac{2\pi (L + D_P)}{\lambda_{GW}}\right)\right] 
\right.
\right.
\nonumber\\
&&~~~~~~~~~~~~~~~~~~~~~~~~~
\left.
\left.
- 
\exp\left[i\left(\frac{2\pi (t_{obs} - L - D_P)}{\lambda_{GW}}\right)\right] 
\right\}
\right]  
\nonumber\\
&&
+ O\left( \frac{h (\lambda_{GW})^2}{D}\right) . 
\label{Res3}
\end{eqnarray}

\section{Beyond the plane-wave approximation}
\subsection{Fresnel and $L/D$ corrections}
In addition to the Fresnel correction in the phase, 
there exist two other corrections. 
One correction comes from the distance difference, 
causing the GW amplitude difference between E and P. 
The fractional difference between paths SE and SP is $O(L/D)$ 
if $L<D$. 
The other correction is due to the difference in GW directions at E and P, 
namely the angle $\phi_{EP}$ between $\bm{N}_E$ and $\bm{N}_P$, 
which satisfies 
\begin{eqnarray}
\sin\phi_{EP} 
&=&
\frac{L \sin\theta}{\sqrt{D^2 + L^2 - 2 D L \cos\theta}} 
\nonumber\\ 
&=& 
\frac{L}{D} \cos\theta 
+O\left( \frac{L^2}{D^2} \right) ,
\label{cos}
\end{eqnarray}
where the cosine formula is used for the triangle EPS 
and the second equality holds only for $L<D$. 
The two next-leading corrections are thus $O(L/D)$.

As an illustration, let us examine the fourth exponential function in Eq. (\ref{Res3}). 
It is expanded in the Fresnel approximation 
\citep{McGrath2021,McGrath2022} 
as 
\begin{eqnarray}
&&
\exp\left[i\left(\frac{2\pi (t_{obs} - L - D_P)}{\lambda_{GW}}\right)\right] 
\nonumber\\
&=& 
\left[
1 + 
\frac{i\pi(1-\cos^2\theta)L^2}{D\lambda_{GW}}
+ 
O\left(
\frac{L^4}{D^2(\lambda_{GW})^2}
\right)
\right]
\nonumber\\
&&
\times
\exp\left[i\left(\frac{2\pi (t_{obs} - D - L (1-\cos\theta))}{\lambda_{GW}}\right)\right] ,
\label{exp}
\end{eqnarray}
where $D_P = (D^2 - 2D L \cos\theta + L^2)^{1/2}$ is expanded in $L/D$. 
The argument of the exponential function in the right-hand side of Eq. (\ref{exp}) 
corresponds to the phase in the plane wave approximation, 
and the second term in front of this function 
can be interpreted as the Fresnel correction 
of $O(L^2/(D\lambda_{GW}))$ 
\citep{McGrath2021,McGrath2022}. 

Therefore, 
the Fresnel correction is still dominant 
in the timing residual also for a fully spherical wavefront, 
whereas 
corrections at $O(L/D)$ are next-leading, 
because $L > \lambda_{GW}$ 
for a typical PTA range. 
These scalings in the timing residual are consistent with those 
for the frequency shift as suggested by 
Eq. (\ref{Doppler2}) and Figure \ref{fig:plot1}.

\subsection{Estimating the scaling of the corrections}
For nearby cases, we make a comparison 
of the amplitude and direction corrections 
at $O(L/D)$ 
to the Fresnel phase correction at $O(L^2/(\lambda_{GW} D))$ 
\citep{Deng2011,McGrath2021,D'Orazio2021,Guo2022}. 
For $\lambda_{GW} \ll L$, 
the former must be smaller than the latter. 
The ratio between them is 
$\sim 0.03 (\lambda_{GW}/30 \mbox{pc})(1 \mbox{kpc}/L)$ 
for a millisecond pulsar at $L \sim 1$ kpc. 
For most of known millisecond pulsars, 
the $L/D$ correction is thus smaller by two or more digits than 
the Fresnel correction.

One may ask if 
corrections of $O(L/D)$ can be comparable to the Fresnel one. 
The nearest millisecond pulsar J0437-4517 
is located at $L = 156$ pc 
\citep{Deller2008}, 
for which the ratio is 
$\sim 0.2 (\lambda_{GW}/30 \mbox{pc})(156 \mbox{pc}/L)$ and 
hence the amplitude and direction corrections are 
comparable to the Fresnel correction.  
For this case, however, all of these corrections are negligible.

\subsection{On nearby GW source candidates}
Once future observations in PTAs detect a GW signal, 
one may ask if the plane-wave ansatz 
is sufficient for the PTA data. 
The corrections at $O(L/D)$ 
are less than roughly $10^{-3}$ 
for $L \sim 10$ kpc and $D > 10$ Mpc, 
for which the distance correction is 
$\sim 0.1$ percents or less. 
Recent PTA bounds on SMBHs within about 500 Mpc 
\citep{Arzoumanian2021a}, most of targeted galaxies are distant 
($> 10$ Mpc). 
However, a few of them are near. 
For instance, J00424433+4116074 is a galaxy at 0.82 Mpc, 
for which 
effects beyond plane waves, especially the Fresnel effect, 
may reach one percent or more. 
In PTA data analysis for galaxies within $D \sim 100$ kpc 
in the local group, 
the effects can be ten percents or more, 
and hence they should be considered.

The corrections can be more important for nearer GW sources.  
The existence of a binary of SMBHs in M31  
is suggested  
\citep{Lauer1993,Bender2005}. 
For such a nearby case, $L/D$ is $\sim 10^{-2}$, 
for which the corrections can be 
at the several percent level.

\subsection{Galactic-center PTA}
There could exist a hidden companion to 
Sagittarius A${}^*$ (Sgr A${}^*$).  
It has been recently 
discussed that the Sgr A${}^*$ observations combined with 
dynamical stability argument seem to rule out 
a $10^5 M_{\odot}$ companion 
\citep{Naoz2020}. 
Even with the companion with $\sim 10^5 M_{\odot}$  
and the orbital radius of $\sim 100$ AU, 
it is expected to be below the typical PTA sensitivity, 
which usually assumes $D_P \sim 10$ kpc for a galactic center source. 

Yet, a large population of pulsars is expected to reside in the galactic center 
\citep{Pfahl}. 
In particular, recent analyses of the gamma-ray emission excess 
using the entire Fermi data 
support that the excess at the galactic center 
can be caused by a population of thousands of undetected millisecond pulsars 
\citep{Ajello,Bartels,Calore,Lee,Gonthier}. 
The first pulsar survey in the galactic center at short millimeter wavelengths, 
using several frequency bands between 84 and 156 GHz, 
has been done, and it has demonstrated that 
surveys at extremely high radio frequencies 
are capable of discovering new pulsars 
\citep{Torne}. 
The survey at a low frequency of $\sim 310$ MHz 
has been also done 
\citep{Hyman}. 

Along this direction, 
an interesting possibility has been argued that PTAs using 
millisecond pulsars within 
0.1-1 pc to Sgr A${}^*$ 
can probe intermediate-mass BHs (IMBHs)    
\citep{Kokcis}, 
where the plane-wave formulae are used.

Let us suppose a hypothetical IMBH with $\sim 10^4 M_{\odot}$ 
orbiting around Sgr A${}^*$ 
with the orbital radius $a \sim 100$ AU, 
for which the typical GW period is $\sim 1$ year. 
For instance, we assume a hypothetical pulsar at 
$\sim 10$ pc from the galactic center. 
This distance is more likely than the speculative value 
of 0.1 pc in \citet{Kokcis}. 
The amplitude of GWs at the position of the pulsar is 
$h \sim m a^2/(D_P (T_{GW})^2) 
\sim Mm/(D_Pa) 
\sim 10^{-14} (M/(10^6 M_{\odot})) 
(m/(10^4 M_{\odot})) (10\, \mbox{pc}/D_P) (100\, \mbox{AU}/a)$, 
where $M$ and $m$ are the mass of Sgr A${}^*$ 
and that of the IMBH, respectively. 

For this hypothetical GW source, 
we shall make an order-of-magnitude estimate 
of the Earth term and the pulsar term in the timing residual. 
By using Eq. (\ref{exp}), these terms are roughly estimated as 
$\mbox{Res}(t_{obs})|_P
\sim 10^{-6} (h_P/10^{-14})(\lambda_{GW}/1\mbox{pc}) \mbox{sec.}$ 
and 
$\mbox{Res}(t_{obs})|_E 
\sim 10^{-9} (h_E/10^{-17})(\lambda_{GW}/1\mbox{pc}) \mbox{sec.}$, 
where 
$h_P$ and $h_E$ denote the GW strain at the pulsar and the Earth, respectively 
and $D$ = 10 kpc, $D_P$ = 10 pc, $\lambda_{GW}$ = 1 pc are assumed. 
The Earth term in the timing residual is smaller by 
$D_P/D \sim 10^{-3}$ than the pulsar term. 

Therefore, 
the Earth term in Eqs. (\ref{Doppler2}) and (\ref{Res2}) 
can be ignored practically 
for the galactic center PTA case that 
$D\sim L \sim 10 \mbox{kpc} \gg D_P  \sim 10 \mbox{pc} 
> \lambda_{GW} \sim 1 \mbox{pc}$, 
for which Eqs. (\ref{Doppler2}) and (\ref{Res2}) are still valid 
since $D_P > L$ is not assumed in the derivation of them. 
Note that expansions in $L/D_P$ 
do not work for this system. 

See also \citet{Guo2022} 
for the detectability of possible nearby GW sources by 
the Square Kilometer Array PTA, 
in which 
the GW direction and amplitude as well as the retarded time 
are considered 
in a fully numerical manner 
based on Eq. (1) of their paper for the frequency shift and 
Eqs. (6) and (7) for the antenna pattern functions.  
Although these equations apparently follow the equations in \citet{Anholm} 
based on plane waveforms, 
integral forms are used in their numerical computations 
for near fields 
\citep{Guo2022}, 
where they do not adopt the far-field approximation.


Finally, 
we mention another potentially useful application. 
It is beam-like GWs 
\citep{Baral2020}, 
for which the wavefront curvature can be significant 
even for a distant GW source. 
However, a generation mechanism of beam-like GWs
is speculative.

\section{Summary}
The frequency-shift and timing-residual formulae were 
derived for GWs with fully spherical wavefronts 
from a compact source. 
We confirmed that the Fresnel correction 
is a leading one under assumptions. 
As a next-leading correction, both 
the GW amplitude and direction corrections 
are at $O(L/D)$. 
A possible relation of nearby GW source candidates 
to the new formula was also mentioned. 
It is left for future to investigate the present formula 
for a larger parameter space in a more general situation. 

\begin{acknowledgments} 
We are grateful to Jolien Creighton, Casey McGrath, 
and Xiao Guo
for the useful comments 
on the earlier version of the manuscript. 
We wish to thank Yuuiti Sendouda and Ryuichi Takahashi for 
fruitful conversations. 
We thank Tatsuya Sasaki and Kohei Yamauchi for 
useful discussions. 
This work was supported 
in part by Japan Society for the Promotion of Science (JSPS) 
Grant-in-Aid for Scientific Research, 
No. 20K03963 (H.A.),  
in part by Ministry of Education, Culture, Sports, Science, and Technology,  
No. 17H06359 (H.A.). 
\end{acknowledgments}




\end{document}